\documentclass[aps,prl,twocolumn,floatix,amssymb,showpacs,amsmath,superscriptaddress]{revtex4-1}
\usepackage{graphicx}
\usepackage{braket}
\usepackage[usenames,dvipsnames]{color}
\usepackage{epsfig,color}
\usepackage{amsmath,amssymb}
\usepackage{mathptmx}
\usepackage[T1]{fontenc}
\usepackage{verbatim}
\usepackage{float}
\usepackage{units}
\usepackage{amsmath}
\usepackage{bm}
\usepackage[colorlinks = true,
            linkcolor = NavyBlue,
            urlcolor  = NavyBlue,
            citecolor = NavyBlue,
            anchorcolor = blue]{hyperref}
\usepackage{epstopdf}
 \usepackage[normalem]{ulem}

\definecolor{orange}{rgb}{1,0.5,0}

\makeatletter

 \@ifundefined{textcolor}{}
 {
   \definecolor{BLACK}{gray}{0}
   \definecolor{WHITE}{gray}{1}
   \definecolor{RED}{rgb}{1,0,0}
   \definecolor{GREEN}{rgb}{0,1,0}
   \definecolor{green4}{cmyk}{1,0,1,0.45}
   \definecolor{BLUE}{rgb}{0,0,1}
   \definecolor{CYAN}{cmyk}{1,0,0,0}
   \definecolor{MAGENTA}{cmyk}{0,1,0,0}
   \definecolor{YELLOW}{cmyk}{0,0,1,0}
 }

\makeatother

\begin{document}

\title{Coupled spin and electron-phonon dynamics at the relativistic Tl/Si(111) surface}
  
\author{Peio Garcia-Goiricelaya}
\affiliation{Materia Kondentsatuaren Fisika Saila, University of the Basque Country UPV/EHU, 48080 Bilbao, Basque Country, Spain}
\affiliation{Donostia International Physics Center (DIPC), Paseo Manuel de Lardizabal 4, 20018 Donostia-San Sebasti\'{a}n, Spain}
\author{Idoia G. Gurtubay}
\affiliation{Materia Kondentsatuaren Fisika Saila, University of the Basque Country UPV/EHU, 48080 Bilbao, Basque Country, Spain}
\affiliation{Donostia International Physics Center (DIPC), Paseo Manuel de Lardizabal 4, 20018 Donostia-San Sebasti\'{a}n, Spain}
\author{Asier Eiguren}
\affiliation{Materia Kondentsatuaren Fisika Saila, University of the Basque Country UPV/EHU, 48080 Bilbao, Basque Country, Spain}
\affiliation{Donostia International Physics Center (DIPC), Paseo Manuel de Lardizabal 4, 20018 Donostia-San Sebasti\'{a}n, Spain}
\date{\today}

\begin{abstract}
  We investigate the role played by the electron spin and the spin-orbit interaction on the exceptional electron-phonon coupling at the Tl/Si(111) surface.
  Our first-principles calculations demonstrate that the particular spin pattern of this system dominates the whole low-energy electron-phonon physics,
  which is remarkably explained by forbidden spin-spin scattering channels.
  In particular, 
  we show that the strength of the electron-phonon coupling 
appears drastically weakened
  for surface states close to the  $\overline{\mathrm{K}}$ and $\overline{\mathrm{K'}}$ 
  valleys, which is unambiguously attributed to the spin polarization
  through the associated modulation due to the spinor overlaps.
  However, close to the $\overline{\Gamma}$ point, the particular spin pattern in this area 
  is less effective in damping the electron-phonon matrix elements, and the result 
  is an exceptional strength of electron-phonon coupling parameter $\lambda \sim 1.4$. 
  These results are rationalized by a simple model for the electron-phonon matrix elements including the spinor terms.
\end{abstract}

\maketitle
%
%
In metals, the charge and spin transport, heat capacity and many other thermodynamical properties, even superconductivity \cite{Grimvall,Mahan},
are strongly influenced by the low-energy dynamics of the so called quasi-particles.
Quasi-particles incorporate the many-body interactions in an approximate way by treating the interacting particles as dressed particles
with modified or renormalized properties \cite{Hedin,Nozieres}.
Electron-phonon dynamics is particularly important in the low-energy domain
and effective masses and transport coefficients are determined mainly by this interaction in clean samples.
Initiated by Migdal \cite{Migdal} and Engelsberg and Schrieffer \cite{EngelsbergSchrieffer} 
the Green's function perturbation theory has been successful in describing realistic properties of materials in combination with 
\textit{ab initio} techniques, even with a high predictive power \cite{Giustinorev}.
The electron-phonon interaction has been investigated systematically in the bulk and 
surface systems,
where it has been established that, as a general rule, the electron-phonon coupling is
enhanced \cite{Plummerep,Hofmannep,Asierprl2002}.
Furthermore, at low dimensional systems, the spin-orbit (SO) coupling 
far from being a mere relativistic  correction, introduces important qualitative changes in the properties of materials,
as the generation of spin-split and spin-polarized electronic states, even in nominally nonmagnetic 
systems \cite{spinpol}.
Likewise, the SO interaction is the responsible of the existence of an entirely new quantum state of matter which is characterized
by its phase-space topology and exhibits exceptional transport properties at the edges
of these materials \cite{TopInsulator}.
Therefore, understanding the low-energy charge and spin coupled dynamics in two-dimensional (2D) systems with strong SO coupling is of capital importance and
a very active research front at the moment \cite{SOCfocus,SOCnat,SOCrashba}.
The coexistence of SO coupling and electron-phonon has been studied considering instructive 
model theoretical treatments combining the Rashba SO term and the Fr\"ohlich/Holstein term, 
revealing an intricate spin dependent polaronic spectral function \cite{polaronmaS0} 
and the impact of the singularity by the Rashba coupling on the electron density of states \cite{ElpheffectsS02D},
among other interesting phenomena associated with the relativistic corrections. 
Examples such as  
the valley dependent electron-phonon coupling of transition-metal dichalchogenides \cite{epsocWS2} and 
the absence of backscattering in topological surface states \cite{epsocBi2Se3} illustrate 
some important aspects related to the spin polarization in presence of the electron-phonon coupling.

In this article, we investigate the electron-phonon dynamics at the Tl/Si(111) surface,
keeping the full spin/momentum dependence of the electron-phonon matrix elements and allowing us to focus on the precise role 
played by the electron spin polarization. 
The Tl/Si(111) shows a very peculiar spin texture \cite{TlSi111julen,TlSi111jon} and 
its band structure combines an in-plane rotational spin for the occupied surface states and a --collinear spin--
valley pattern for the unoccupied ones. 
Therefore, the geometry of the electron-phonon coupling is radically different for the occupied and unoccupied spin splitted surface states,
but both energy regions should be accessible under doping.
We analyze the energy and momentum dependence of the mass enhancement parameter for low energy surface states, below 
and above the Fermi level, and we find a remarkable strength of the electron-phonon coupling.
Moreover, in order to clarify the role of the spin polarization, we fix the electron momentum $\mathbf{k}$ at some representative points,
and break up the $\mathbf{q}$ dependent contributions from the entire Brillouin zone, which allows us to 
correlate the results with the spin pattern of this surface.
An extreme  example illustrating how the spin structure enters the electron-phonon physics, occurs when a
phonon mode with momentum $\mathbf{q}$ connects two points of the surface Brillouin zone (SBZ)
with orthogonal spin-states causing 
an almost perfect extinction of the electron-phonon scattering. This picture
applies perfectly for the unoccupied part of the band structure, as this area shows an ideal 
valley arrangement. However, the 
spin exclusion mechanism described above is not enough for weakening the exceptional  magnitude 
of the electron-phonon interaction in the case of the occupied surface bands.
%

%
The Tl/Si(111) surface was simulated considering a slab consisting of 10 silicon layers with a thallium monolayer on one termination of the slab.
The silicon dangling bonds at the other end of the slab were saturated inserting a single hydrogen coverage as in Ref. \cite{TlSi111julen,TlSi111jon}.
First-principles computations were performed using noncollinear DFT \cite{Hohenberg-Kohn,Sham-Kohn} and DFPT \cite{DFPT} approaches with fully relativistic
norm-conserving pseudo-potentials \cite{NCP} as implemented in the \textsc{Quantum Espresso} package \cite{SOCprb,SOCdalcorso,QE}
and using the PBE-GGA parametrization \cite{GGA} for the exchange-correlation energy.
All atomic forces were relaxed up to at least $10^{-5} \mathrm{Ry/a.u.}$.
We used a $24\times24$ Monkhorst-Pack grid for self-consistent electronic calculations, while phonon modes were calculated in a coarse $8\times8$ \textit{q}-point mesh.
The SBZ integrals involved in the computation of the electron-phonon coupling were evaluated in very dense (10$^6$) \textit{k-} and \textit{q}-grids 
by means of a Wannier interpolation scheme \cite{Giustino-prl2007,asier-W}.
Electron-phonon matrix elements were calculated considering noncollinear spinor wave functions.

\begin{figure}[ht!]
 \begin{center}
  \includegraphics[width=1\columnwidth,angle=0,scale=1.0]{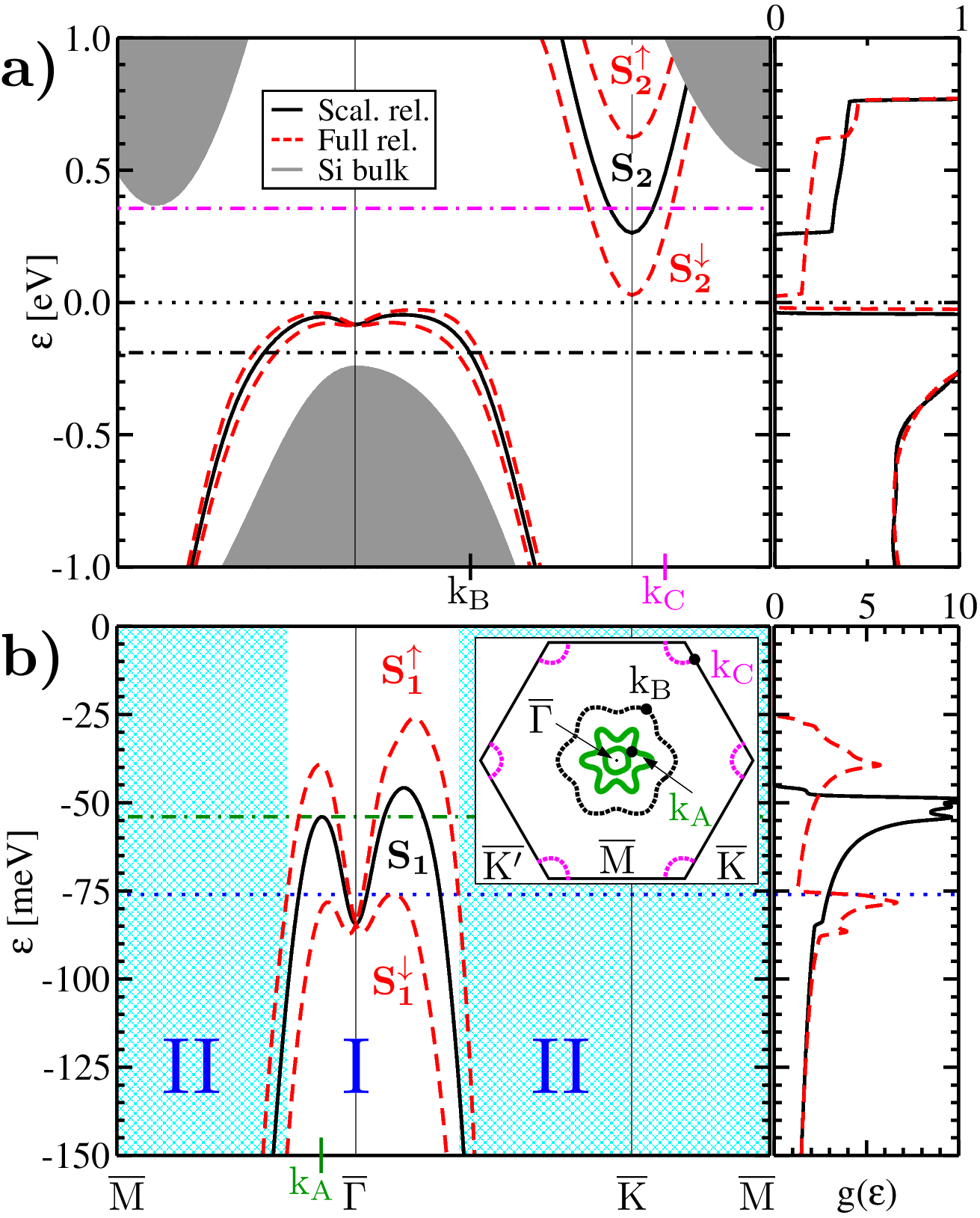}
 \end{center}
 \caption{Electron band structure for the eV (a) and meV (b) range of energy (left) and the corresponding DOS (right),
of the Tl/Si(111) surface.
          The scalar and fully relativistic calculations are represented by black solid and red dashed lines, respectively.
          The gray shaded area in (a) is the bulk band projection.
          The inset in panel (b) shows the energy contour of the highest occupied band at
          $\mathbf{k}=\mathrm{k_A}$ and $\mathbf{k}=\mathrm{k_B}$ and
          the lowest unoccupied band at $\mathbf{k}=\mathrm{k_C}$ in the scalar case.
          The corresponding energy levels are plotted by dashed dotted-lines in (a) for $\mathrm{k_B}$ (black) and $\mathrm{k_C}$ (magenta) and in (b) for $\mathrm{k_A}$ (green).} 
 \label{fig:Fig1}
\end{figure}

%
Figure\,\ref{fig:Fig1} shows the electronic structure of the Tl/Si(111) surface.
The black solid  and red dashed lines show the scalar (i.e. without SO coupling) and fully (i.e. spin-split) relativistic  
surface state results, respectively. 
The left-hand side of panels (a) and (b) display the electron surface band structure with respect to the Fermi level (dotted black line in (a)) in the eV and meV range,
respectively, while those on the right-hand side represent the corresponding density of states (DOS).
The silicon bulk band projection is illustrated by the grey shaded area.
These calculations show an excellent agreement with photo-emission measurements \cite{TlSi111sakamotonat,TlSi111sakamotoprl,TlSi111stolwijk}
and previous theoretical studies \cite{TlSi111julen,TlSi111jon}.
When SO effects are included, the highest occupied S$_1$ scalar surface band spin-splits into the S$_{1}^{\uparrow}$ and S$_{1}^{\downarrow}$ bands,
dominating the low-energy  region close to the Fermi level. 
The lowest unoccupied S$_2$ surface band, on the other hand, spin-splits into S$_{2}^{\uparrow}$ and S$_{2}^{\downarrow}$,
yielding the strongest spin-split energy known in literature ($\sim$ 0.6 eV at $\overline{\rm{K}}$).
The inset of Fig.\,\ref{fig:Fig1}(b) shows the constant energy surfaces/contours corresponding to the scalar energies given by the dashed-dotted lines for three selected
carrier momenta  in the band structure plot: $\mathrm{k_A}$ and $\mathrm{k_B}$ for the occupied $\rm S_1$ surface band, and $\mathrm{k_C}$ for the unoccupied S$_2$ band.
While the S$_{2}$ surface band appears isotropic around $\overline{\mathrm{K}}$  and $\overline{\mathrm{K'}}$,
the contour corresponding to the S$_{1}$ surface band presents a hexagonal -daisy flower like- anisotropy.
The same holds for the spin-split S$_{2}^{\uparrow,\downarrow}$ and S$_{1}^{\uparrow,\downarrow}$ surface bands, respectively,
except that the latter will show a double concentric structure, as shown in the right-hand panel of Fig.\,\ref{fig:Fig4}(c).
The  DOS corresponding to the scalar S$_{1}$ surface band exhibits a single 1D-like singularity at the top of the band.
For the spin polarized surface bands (S$_{1}^{\uparrow, \downarrow}$),
each state shows this singularity.
Obviously, the electron-phonon coupling is enhanced close to the phase-space connected to these singularities, as it will be demonstrated shortly.
\begin{figure}[ht!]
 \begin{center}
  \includegraphics[width=1\columnwidth,angle=0,scale=1.0]{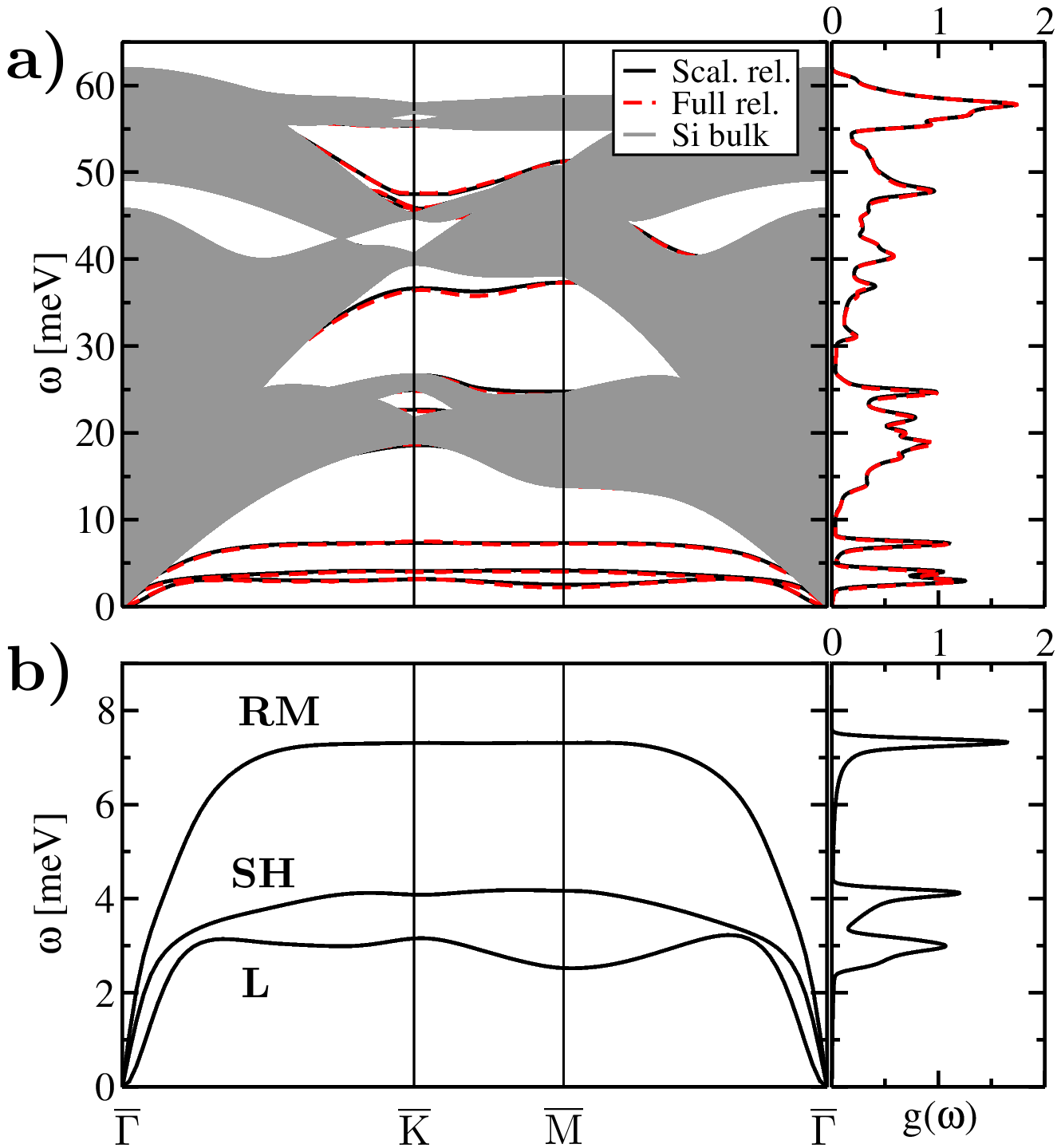}
 \end{center}
 \caption{(a) Phonon dispersion relation (left)  and corresponding PHDOS (right)  of the Tl/Si(111) surface.
         (b)  Vibrational spectrum for the three acoustic modes. The scalar and fully relativistic calculations are represented by solid black and dashed red lines, respectively.
          The shaded grey area in (a) is the bulk band projection.}
 \label{fig:Fig2}
\end{figure}

%
Figure\,\ref{fig:Fig2} displays the phonon dispersion at the Tl/Si(111) surface for the complete energy range (a) and the 0-9 meV range (b),
the latter showing only the three low-energy acoustic surface phonon modes, which correspond fundamentally to 
localized vibrations connected to Tl displacements.
Ordered by increasing energy, we label these modes according to their 
polarization near $\overline\Gamma$, 
which correspond essentially to the longitudinal (L), shear horizontal (SH) and the Rayleigh mode (RM).
These will be the most relevant modes for the coupling with the electron surface states.
%
%
We proceed now to analyze the electronic energy ($E_{F}=\varepsilon_{i}^{\mathbf{k}}$) and momentum ($\mathbf{k}$)
dependence of the electron-phonon mass enhancement parameter ($\lambda$) defined as,
\begin{equation}
\lambda_{\mathbf{k},i}=\sum_{\mathbf{k'}j\nu}
\frac{|g_{ij}^{\nu}(\mathbf{k},\mathbf{k'})|^{2}}
		{\omega^{\mathbf{k}-\mathbf{k'}}_{\nu}}
 \delta(\varepsilon_{i}^{\mathbf{k}}-\varepsilon_{j}^{\mathbf{k'}}\pm\omega_{\nu}^{\mathbf{k}-\mathbf{k'}})
\label{eq:lk}.
\end{equation}
In Eq.\,\ref{eq:lk}, $g_{ij}^{\nu}(\mathbf{k},\mathbf{k'})$ denote the electron-phonon matrix elements,
$\varepsilon_{i}^{\mathbf{k}}$ are the bare electron band energies
and 
$\omega^{\mathbf{k}-\mathbf{k'}}_{\nu}$ represent the phonon energies for 
momentum $\mathbf{q}=\mathbf{k}-\mathbf{k'}$.
The $\lambda$ parameter is one of the most representative magnitudes of the strength of the electron-phonon coupling and, strictly speaking,
is a physical quantity defined exactly at the Fermi level ($E_{F}$) \cite{Grimvall}.
Therefore, Eq.\,\ref{eq:lk} should be understood as a Fermi level varying analysis ($E_{F}=\varepsilon_{i}^{\mathbf{k}}$).
\begin{figure}[ht!]
 \begin{center}
  \includegraphics[width=1\columnwidth,angle=0,scale=1.0]{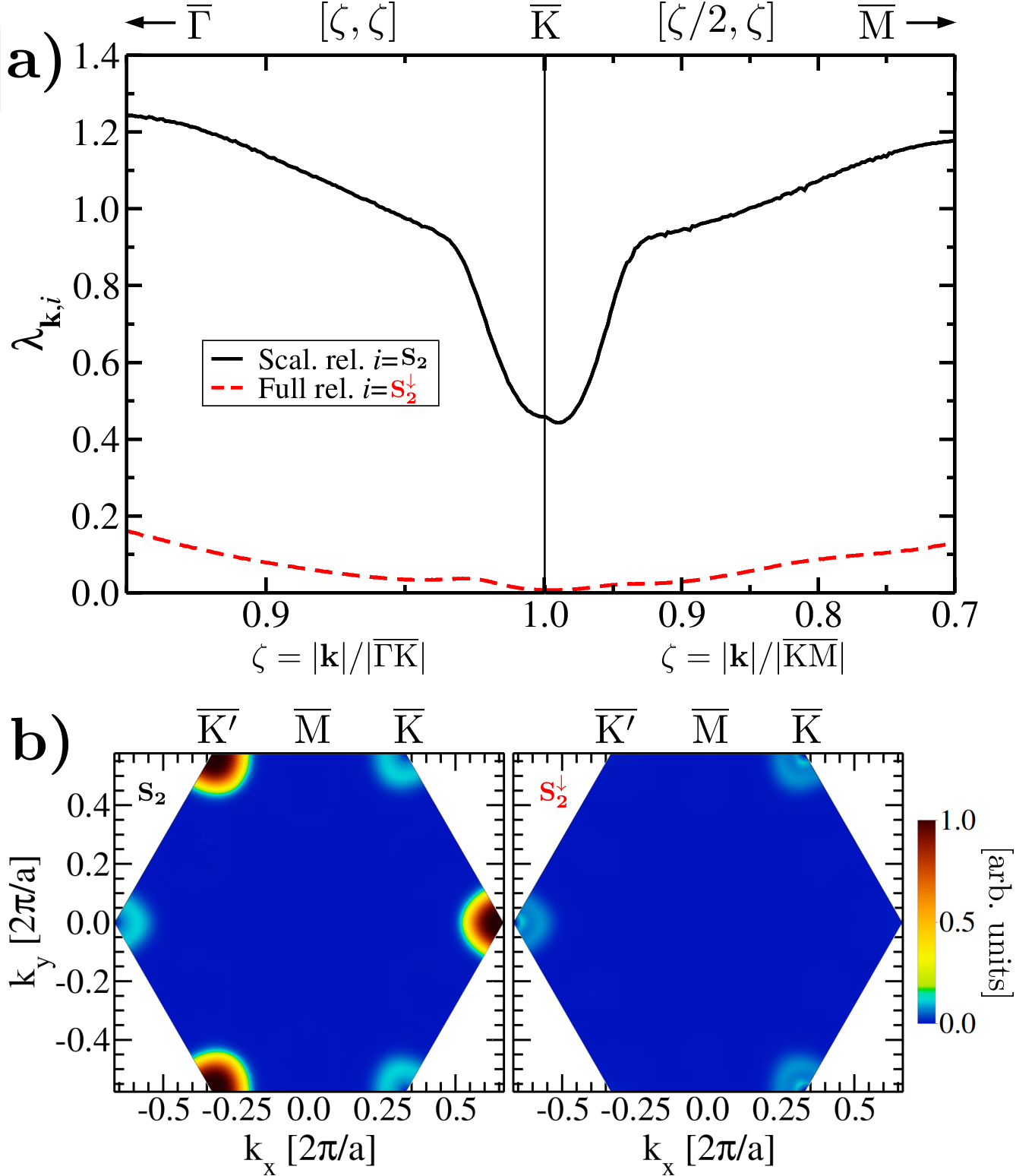}
 \end{center}
 \caption{(a) Momentum-dependent electron-phonon $\lambda_{\mathbf{k}, i}$ parameter for the lowest unoccupied band around $\overline{\mathrm{K}}$
          for the scalar relativistic S$_{2}$ surface band (solid black) and for the spin-split S$_{2}^{\downarrow}$ surface band (dashed red).
          (b) ${\mathbf{k'}}$ momentum resolved 
	contributions to $\lambda_{{\mathbf{k}={\overline{\mathrm K}}},i}$ (see Eq.\,~\ref{eq:lk})
          within the SBZ for the scalar S$_{2}$ (left) and spin-split S$_2^{\downarrow}$ (right) surface bands.
} 
 \label{fig:Fig3}
\end{figure}
%
%
Figure\,\ref{fig:Fig3}(a) shows the calculated $\lambda_{\mathbf{k},i}$ for the unoccupied scalar S$_2$ surface band (solid black)
and for the spin-split S$_{2}^{\downarrow}$ surface band (dashed red),
as a function of the electron momentum around the valley $\overline{\rm{K}}$, and along the high symmetry directions $\overline{\Gamma\mathrm{K}}$ and $\overline{\mathrm{K}\mathrm{M}}$
\footnote{For the spin-split S$_{2}^{\uparrow}$ band, the coupling with bulk states dominates and it is outside the scope of this work.}.
As mentioned before, the three low-energy phonon modes contribute to 98\% of the coupling strength.
The quantitative and qualitative differences between the scalar and fully relativistic results are striking. 
At $\overline{\mathrm{K}}$, $\lambda_{\overline{\mathrm{K}},\mathrm{S}_{2}^{\downarrow}}$ vanishes for the spin-split band,
which is in complete contrast with the scalar case, where we find  $\lambda_{\overline{\mathrm{K}},\mathrm{S_2}}=0.45$.
Moving away from the center of the valley $\overline{\mathrm{K}}$, our calculations show that the momentum-resolved mass enhancement parameter 
is about one order of magnitude stronger for the scalar relativistic band. 
The reason for the step-like behavior of the latter is that, in the immediate vicinity of the bottom of the valley, 
no phonon-emitting electron scattering channel is energetically accessible and only the hole-phonon processes are possible,
which means that solely the term $\delta(\varepsilon_{i}^{\mathbf{k}}-\varepsilon_{j}^{\mathbf{k'}}+\omega_{\nu}^{\mathbf{k}-\mathbf{k'}})$ contributes to Eq.\,\ref{eq:lk}. 
However, when the difference between the energy $\varepsilon_{i}^{\mathbf{k}}$ 
 and that of the bottom of the valley equals,
or  is greater than the smallest surface phonon energy, phonon-emitting electron scattering events ($\delta(\varepsilon_{i}^{\mathbf{k}}-\varepsilon_{j}^{\mathbf{k'}}-\omega_{\nu}^{\mathbf{k}-\mathbf{k'}})$)
are also allowed and $\lambda_{\mathbf{k},\mathrm{S}_2}$ reaches  a value of 0.9.
As momentum/energy increases, the coupling strength grows smoothly following the same trend as the DOS in Fig.\,\ref{fig:Fig1}(a).
For the spin-split S$_{2}^{\downarrow}$ surface states, $\lambda_{\mathbf{k},\mathrm{S}_{2}^{\downarrow}}$ is practically zero close to the bottom of the valley
and, as energy increases, it grows up to values not larger than 0.2.
Fig.\,\ref{fig:Fig3}(b) helps to understand the difference between the spin-split and scalar results.
The left (right) panel shows the contributions from each point in the SBZ to the calculation of $\lambda_{\mathbf{k},i}$ by means of Eq.\,\ref{eq:lk}
for the scalar S$_2$ (spin-split S$_{2}^{\downarrow}$) case when the electron momentum is fixed to be $\mathbf{k}=\mathrm{\overline K}$.
The obvious difference appears at $\mathbf{k'}=\overline{\mathrm{K'}}$, where the contribution is maximum in the scalar relativistic case
but results to be negligible in the fully relativistic one.
Actually, spin-split unoccupied surface spinor states have surface-perpendicular opposite spin polarizations at $\overline{\mathrm{K}}$ and $\overline{\mathrm{K'}}$ 
\cite{TlSi111julen,TlSi111jon,TlSi111sakamotonat,TlSi111sakamotoprl,TlSi111stolwijk}, and therefore inter-valley ($\overline{\mathrm{K}}\rightarrow\overline{\mathrm{K'}}$) scattering via
phonon-emission is forbidden, which is in complete contrast with the scalar case where this scattering channel is perfectly accessible.
Intra-valley ($\overline{\mathrm{K}}\rightarrow\overline{\mathrm{K}}$ or $\overline{\mathrm{K'}}\rightarrow\overline{\mathrm{K'}}$) decay channels, 
on the other hand, are allowed in both scalar and fully relativistic cases, although with a smaller probability (see the color scale in Fig.\,\ref{fig:Fig3}(b)).
\begin{figure}[ht!]
 \begin{center}
  \includegraphics[width=1\columnwidth,angle=0,scale=1.0]{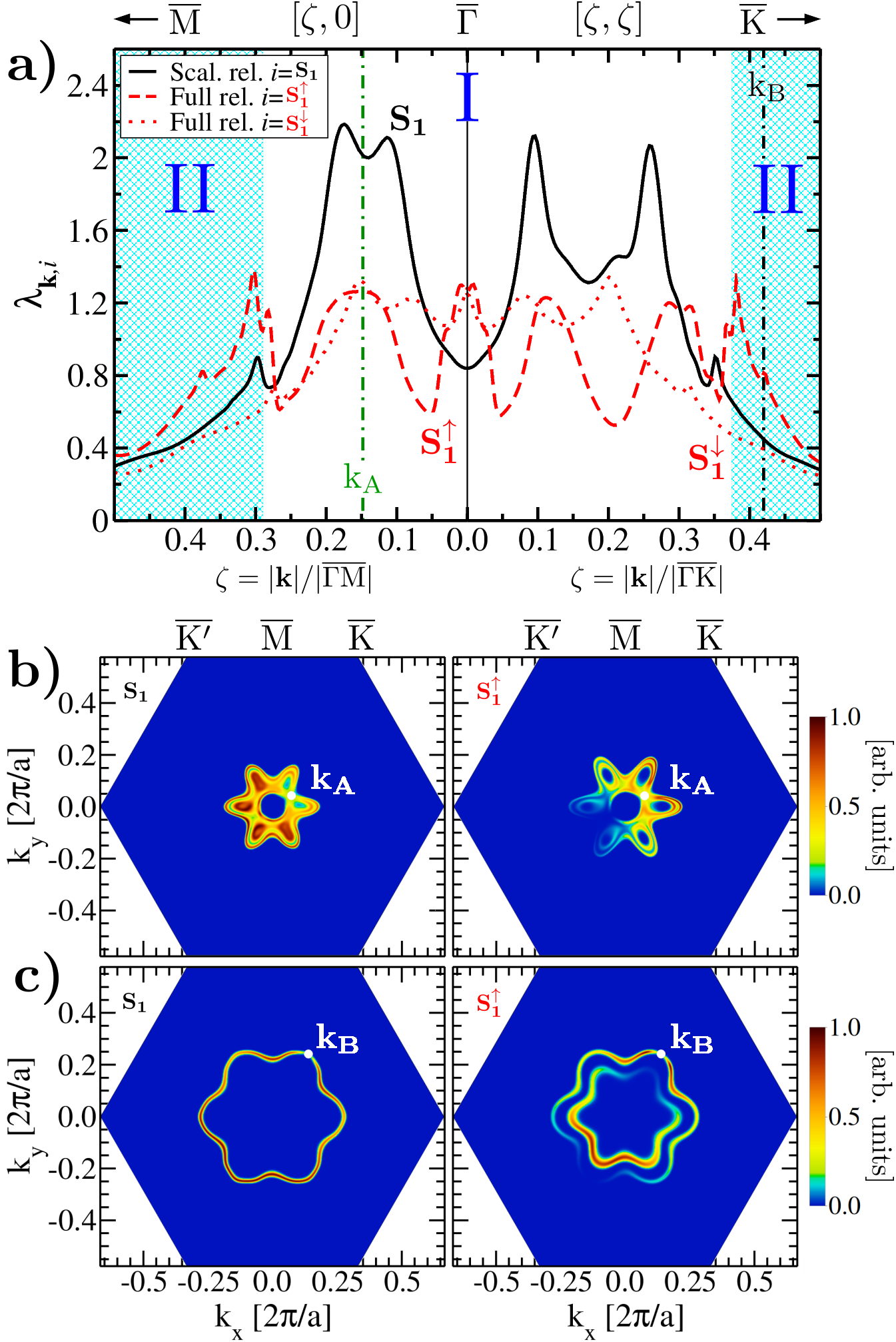}
 \end{center}
 \caption{(a) Momentum-dependent electron-phonon $\lambda_{\mathbf{k}, i}$ parameter for the highest occupied bands around $\overline{\Gamma}$.
The solid black line displays the result for the  scalar S$_{1}$ surface band while dashed and dotted red lines show the spin-split
 S$_{1}^{\uparrow}$ and S$_{1}^{\downarrow}$ results, respectively.
     Regions I and II correspond to the ones in Fig.\,\ref{fig:Fig1}.
Panels (b) and (c) show the 
          ${\mathbf{k'}}$ momentum resolved contributions to
 $\lambda_{\mathbf{k}, i}$ 
(see Eq.\,~\ref{eq:lk})
within the SBZ for the scalar S$_{1}$
 (left) and spin-split S$_1^{\uparrow}$ (right) surface bands
when $\mathbf{k}=\mathrm{k_A}$ (region I) and  $\mathbf{k}=\mathrm{k_B}$ (region II),
respectively. $\mathrm{k_A}$ and $\mathrm{k_B}$ are shown in Fig.\,\ref{fig:Fig1}}
 \label{fig:Fig4}
\end{figure}

%
Figure\,\ref{fig:Fig4}(a) shows the momentum dependence of the $\lambda$ parameter for the occupied surface states close to $\overline{\Gamma}$
in the scalar S$_1$ band (solid black), and in the spin-split S$_{1}^{\downarrow}$ (dotted red) and S$_{1}^{\uparrow}$ 
(dashed red) bands, along the high symmetry directions $\overline{\Gamma\mathrm{M}}$ and $\overline{\Gamma\mathrm{K}}$.  
In this case 95\% of the contribution comes from the three acoustic phonon modes.
We can distinguish two different regimes in Fig.\,\ref{fig:Fig4}(a):
In region II, 
the scalar and fully relativistic results 
are similar, which is in contrast with region I,
where even if 
the fully relativistic electron-phonon coupling strengths are damped by up to a 
factor of 3 compared to the scalar results, we still find values of $\lambda$ as large as  $1.4$
\footnote{A simple McMillan formula estimate gives a superconducting transition temperature of $T_c\approx 7 $K for the relativistic case but a strong coupling analysis is in order.}.
The phase space of the regions I and II is also displayed in the 
patterned blue areas of the surface band structure of 
Fig.\,\ref{fig:Fig1}(b):
region I above the horizontal blue dotted line corresponds to the phase space of the spin-split S$_{1}^{\uparrow}$ surface states where only intra-band scattering processes are allowed,
while below this line and in region II, intra-band 
 or and inter-band processes are accessible for both S$_{1}^{\uparrow}$ and S$_{1}^{\downarrow}$.
Fig.\,\ref{fig:Fig4}(b) shows the momentum resolved contributions to
 $\lambda_{\mathbf{k},i}$ (Eq.\,\ref{eq:lk})
for the scalar S$_1$ (left) and spin-split S$_{1}^{\uparrow}$ (right)
 surface bands for  $\mathbf{k}=\mathrm{k_A}$ belonging
to region I (see Fig.\,\ref{fig:Fig1}(b)).
Fig.\,\ref{fig:Fig4}(c) gives the corresponding results for $\mathbf{k}=\mathrm{k_B}$ within region II.
Clearly, the contributions shown in the left and right panels of Fig.\,\ref{fig:Fig4}(c) are very similar,
whereas Fig.\,\ref{fig:Fig4}(b) shows that they are strongly weakened when SO interaction is included.
This is easily understood in terms of spin state overlaps.
Let us recall that occupied spin-split surface bands have a Rashba-like \cite{Rashba-Bychkov} chiral spin polarization \cite{TlSi111julen,TlSi111jon} on the surface plane near $\overline{\Gamma}$.
In the Rashba model, the spinor overlaps 
 between two different states within the same spin-split
 band (intra-band) appear modulated appropriately by $(1+\cos\theta)/2$,
$\theta$ being the angle between
the initial and final momentum of the electron.
Yet, when the overlap happens between two states belonging to different spin-split bands (interband),
overlaps vary
  as $(1-\cos\theta)/2$, since the spin
is polarized in opposite directions in each band. 
Therefore, in region II, where both inter-band and intra-band
channels are allowed, the results are qualitatively similar 
to the spin-degenerate case (Fig.\,\ref{fig:Fig4}(c)).
However, in region I, where only intra-band contributions are accessible for the S$_{1}^{\uparrow}$ band, the matrix elements are reduced by
$(1+\cos\theta)/2$, as  shown in Fig.\,\ref{fig:Fig4}(b).
In the case of  S$_{1}^{\downarrow}$, although both intra and inter-band scattering processes are allowed,
$\lambda_{\mathbf{k},\mathrm{S}_{1}^{\downarrow}}$ still shows a noticeable attenuation in
 region I compared to the scalar case.
This is attributed to the difference between the scalar and spin-split electronic density of states.
%

%
In summary, we demonstrate the fundamental role played by the electron spin and the relativistic effects on the 
dynamics of surface electrons at the Tl/Si(111) surface.
We have calculated the state-dependent electron-phonon
mass enhancement parameter and shown that the spin polarization 
of the surface states enters in a decisive way by modulating the electron-phonon 
matrix elements which we have demonstrated unambiguously by representing the -squared- matrix elements
though the entire Brillouin zone.
We show that the electron-phonon coupling appears strongly weakened for the unocupied bands (S$_2$) in this surface, 
as these bands are arranged in a collinear spin valley structure.
More interestingly, while the restriction imposed by the spin polarization 
should also apply to the occupied surface bands (S$_1$), the strength of the 
coupling remains remarkably high ($\lambda \sim 1.4$) which will require a deeper understanding 
of transport and superconducting properties in this surface. 
We believe that this work should stimulate further experimental and theoretical investigation in this research front.
%

%
The authors are grateful to I. Etxebarria and J. Iba\~{n}ez-Azpiroz for fruitful discussions,
and acknowledge the Department of Education, Universities and Research of the Basque Government and UPV/EHU (Grant No. IT756-13)
and the Spanish Ministry of Economy and Competitiveness MINECO (Grant No. FIS2013-48286-C2-1-P and FIS2016-75862-P) for financial support.
Computer facilities  were provided by the Donostia International Physics Center (DIPC).
P.G. acknowledges financial support from UPV/EHU (Grant. No PIF/UPV/12/279) and DIPC.
%


%
\end{document}